\documentclass[aps,prl,10pt,reprint,groupedaddress,superscriptaddress]{revtex4-2}

\usepackage{graphicx}
\usepackage{dcolumn}
\usepackage{bm}
\usepackage[nointegrals]{wasysym}
\usepackage{hyperref}
\usepackage{graphicx}
\usepackage{epstopdf}
\usepackage{color}
\usepackage{amsmath}
\usepackage{amsfonts}
\usepackage[normalem]{ulem}
\DeclareMathOperator{\Tr}{Tr}
\usepackage{braket}
\usepackage[T1]{fontenc}
\definecolor{orange}{RGB}{255,127,0}
\definecolor{blue2}{RGB}{33,114,173}
\DeclareMathOperator{\sgn}{sgn}

\usepackage{pdfpages} 
\usepackage{pgffor} 

\makeatletter
\AtBeginDocument{\let\LS@rot\@undefined}
\makeatother

\def\supplementfilename{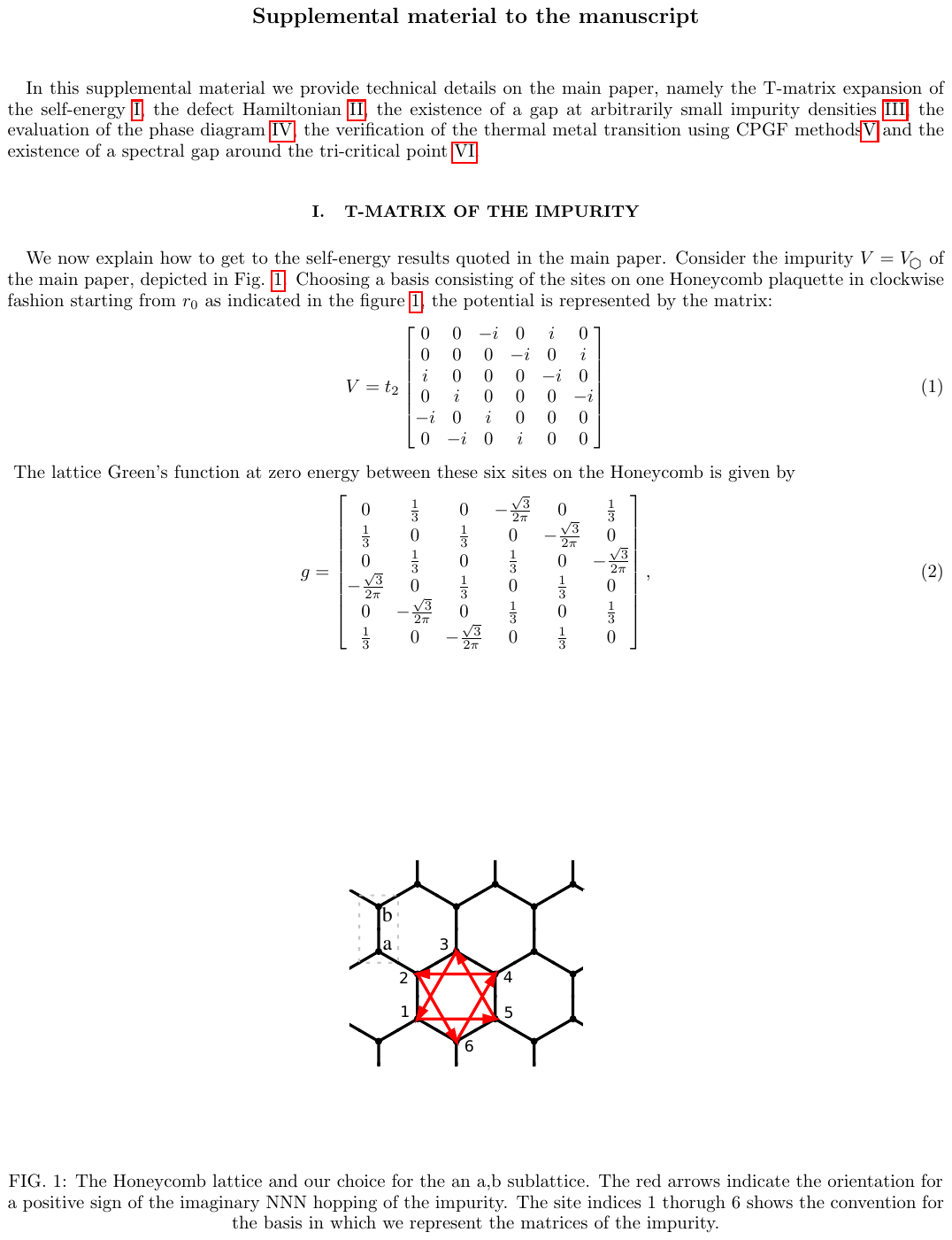}

\pdfximage{\supplementfilename}
\def\numbersupplementpages{\the\pdflastximagepages}

\newif\ifarXiv
\arXivtrue 

\begin{document}
\preprint{APS/123-QED}

\title{Genuine topological Anderson insulator from impurity induced chirality reversal}

\newcommand{\TUM}{\affiliation{Technical University of Munich, TUM School of Natural Sciences, Physics Department, 85748 Garching, Germany}}
\newcommand{\MCQST}{\affiliation{Munich Center for Quantum Science and Technology (MCQST), Schellingstr. 4, 80799 M{\"u}nchen, Germany}}
\newcommand{\Imperial}{\affiliation{Blackett Laboratory, Imperial College London, London SW7 2AZ, United Kingdom}}

\author{Avedis Neehus} \TUM \MCQST
\author{Frank Pollmann} \TUM \MCQST
\author{Johannes Knolle} \TUM \MCQST \Imperial

\date{June 2, 2025}%

\begin{abstract}
We investigate a model of Dirac fermions with  mass impurities which open a global topological gap even in the dilute limit. Surprisingly, we find that the chirality of this mass term, i.e., the sign of the Chern number, can be reversed by tuning the magnitude of the single-impurity scattering. 
Consequently, the disorder induces a phase disconnected from the simple clean limit of the topological insulator, which is achieved via an impurity resonance inducing additional bands with Chern number two. Thus, we call this impurity induced phase a genuine topological Anderson insulator. In seeming contradiction to the expectation that mass disorder is an irrelevant perturbation to the Dirac semi-metal, the tri-critical point separating  these two Chern insulating phases and a thermal metal phase is located at zero impurity density and connected to the appearance of a zero energy bound state in the continuum that corresponds  to a perfectly resonant mass impurity. Our conclusions based on the T-matrix expansion are substantiated by large scale Chebyshev-Polynomial-Green-Function numerics. We discuss possible experimental platforms. 
\end{abstract}

\maketitle

\emph{Introduction.}---Topological phases of matter \cite{moessner_moore_2021}  are not merely robust to disorder, but rather disorder is an integral part of them. This relationship is prominently exhibited in experiments on the quantum Hall effect (QHE) \cite{1986Klitzing, Klitzing1980} where the hallmark signature of the QHE, the resistivity plateaus, may only be observed in a dirty sample where disorder broadens the Landau levels and allows for a continuous dependence of the Fermi-level on the magnetic field \cite{1994Bellisard}. Furthermore, the edge-modes corresponding to these plateaus \cite{Rhim18} are themselves defined through their lack of back-scattering with impurities. 

Nevertheless, disorder is usually seen as inimical to topological phases because it tends to localize the extended states that support them \cite{Anderson58}.
With the discovery of the topological Anderson insulator (TAI) \cite{Li2009,Groth_2009,meier2018observation}, it appeared that there was a topological phase existing not in spite but because of disorder. Soon after, however, it was found that the disorder merely enlarges the topologically non-trivial parameter space of the clean model, in other words the disorder can be  turned off continuously~\cite{Prodan11,lieu2018disorder}. 

Another phenomenon where disorder tends to delocalize rather than localize is that of Anderson anti-localization; in certain symmetry classes \cite{Altland_1997} the quantum corrections to the Drude model can drive the system into a disorder induced \emph{metallic} phase \cite{Evers2008}, e.g.  systems in symmetry class $D$ with only particle-hole symmetry (PH) \cite{Bocquet_2000}. Here the metallic phase described by a \emph{perfect metal} fixed point is called a thermal metal (TM) alluding to the fact that in the commonly discussed realizations, e.g. $p$-wave superconductors, $\nu=5/2$ fractional QH states \cite{Senthil2000,Read2000a, Laumann2012} or Kitaev quantum spin liquids (QSL) \cite{Nasu2017, Self2019},  the quasi-particles carry only heat currents as particle number conservation and spin-rotation invariance are broken. Because in 2D (anti-)localization corrections are marginal, numerical studies of Anderson transitions are notoriously complicated. Indeed, the body of works \cite{Cho97,Chalker01, Read2000b, Gruzberg01, Merz02, Merz202, Mildenberger06, Mildenberger07, Kagalovsk08, Kagalovsky2010, Beenakker10, Lian18,Pan_2021, Wang21} relying on network models remains inconclusive as to the structure of the RG-flow in class $D$ systems, the main point of content being the nature and position of the tri-critical point that separates (Chern) insulators and the thermal metal phase. What is generically agreed upon is that the free fixed point controlling the clean integer  QH transition is stable to weak mass disorder and the tri-critical point must be located at a finite disorder strength.  

\begin{figure}
    \includegraphics[width=0.47\textwidth]{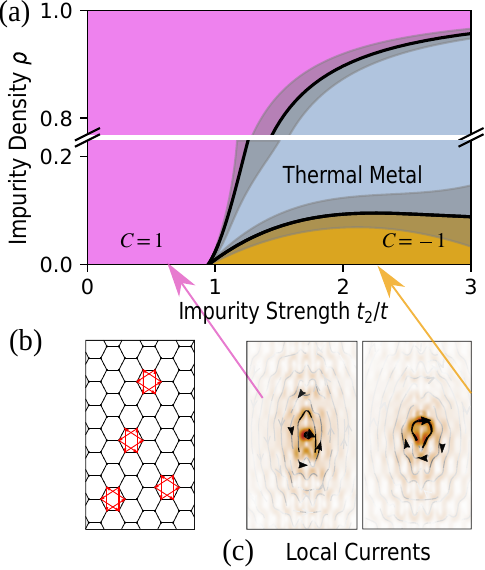}
    \caption{\textbf{(a)} Phase diagram of the nearest-neighbour honeycomb model with topological mass disorder, which is depicted in panel \textbf{(b)}. The red arrows represent the imaginary hoppings of the impurity potential. \textbf{(c)} Streamplot of the local current  vector field around a single impurity. The orange hue is proportional to the magnitude of the local vector field. The probability currents are evaluated on a state with energy close to 0, inside the $C = 1$ (left) and the $C = -1$ phase (right).}
    \label{fig:enter-label}
\end{figure}

In this work we study a disorder potential $V$, which takes the form of a topological mass impurity, and uncover hitherto unexplored aspects of the interplay between disorder and topology: Firstly, we find a transition between different CI phases mediated by impurity resonances whose hybridization with the Dirac cone leads to the emergence of \emph{new topological mobility gaps} that have no clean equivalent \footnote{In this case the clean system refers to the Haldane model, i.e.  an impurity density of one.}. We dub this phase that is enabled by a Chernful impurity resonance band a genuine TAI (GTAI).  Secondly, we show that our model realizes the tri-critical point of class $D$ systems at \emph{zero impurity density}, i.e., the clean Dirac fermions seem unstable to mass disorder, see Fig. \ref{fig:enter-label} (a). 
For weak scatterers we observe that $V$ opens a gap with Chern number $C = 1$ at \emph{any} finite density. Since the Haldane model is recovered in the limit of the impurity density going to one, this does not represent  the GTAI. While the Haldane model itself is stable against weak perturbations, we find that for strong enough $V$ there is a region of densities which hosts a thermal metal phase. However, for strong dilute impurities there appears yet again a Chern insulating phase, which is a GTAI  with the effective chirality of the impurity potential \emph{reversed} such that $C = -1$, see Fig. \ref{fig:enter-label} (c). At the tri-critical point between the two Chern phases and the TM we find a zero energy bound state (ZBS) which we could not associate with known mechanisms \cite{Hsu2016,Teo10}. Strikingly, the ZBS corresponds to an impurity whose  renormalized potential is a divergent Dirac \emph{mass} term. 
We discuss these results both within the T-matrix approximation and state-of-the-art exact numerics with up to $O(10^9)$ sites. 

\emph{The model.}---  We consider a nearest neighbour (NN) tight-binding model of spin polarized fermions $c$ on a hexagonal lattice together with an impurity potential that takes the form of the imaginary next-NN hopping, Fig. \ref{fig:enter-label} (b):
\begin{align}
    H(t_2, \rho)&=-t \sum_{\langle i, j\rangle} c_i^{\dagger} c_j+ \sum_{\varhexagon} p(\varhexagon, \rho) V_{\varhexagon}(t_2) \\
    V_{\varhexagon}  &=  t_2 \sum_{\langle\langle i, j\rangle\rangle \in \hexagon}  e^{i \Phi_{i,j}} c_i^{\dagger} c_j.
\end{align}
It is distinguished from the regular Haldane model \cite{Haldane1988} by introduction of the  random binary variable $p$ which is equal to 1 on a fraction $\rho$ of all the available honeycombs and 0 otherwise. We let $t_2 >0$ and take $|{\Phi}| = \frac{\pi}{2}$ to study the rich phase diagram of class $D$ systems and keep the impurity potential simple \footnote{With a finite real part we would introduce an additional chemical potential.}. However, the existence of the GTAI is generic for a finite range of $\Phi$.

\emph{T-matrix Approximation.}--- First we study the system perturbatively and introduce $T$  as the renormalized potential $V$ taking into account multiple scattering events off of a single impurity. It is computed by \cite{TAKENO1969641}
\begin{align}
       T_{\varhexagon} (E) = (1- V_{\varhexagon}G_0(E))^{-1}V_{\varhexagon}, 
\end{align}
where $G_0(E)$ is the lattice Greens function of the clean system. Like the potential $V$, $T$ is represented by a $6 \times 6$ matrix which we calculate exactly. Following Ostrovsky \emph{et al.} \cite{Kot2020, Ostrovsky2010, Schelter2011} we project $T$ into the basis of the two-flavor Dirac spinors in terms of which we may write the low-energy theory $H_{\boldsymbol{R}} = \int d\boldsymbol{r} \psi^\dagger (\boldsymbol{r}) h_{\boldsymbol{R}}(\boldsymbol{r}) \psi(\boldsymbol{r})$ of the model with an impurity at a \emph{certain} position $\boldsymbol{R}$ as 
\begin{align}
    h_{\boldsymbol{R}}(\boldsymbol{r}) = \frac{3t}{2}&\left( \tau_z \sigma_x i\partial_x+\sigma_y i\partial_y \right) \label{eq:clean}+\\  \delta_a(\boldsymbol{r}-\boldsymbol{R}) \frac{3t_2}{2}&\left(- \tau_x \sigma_z i\partial_x+\tau_y i\partial_y +\sigma_z\tau_z \right). \nonumber 
\end{align}
Here the Paulis $\sigma$ act on the sublattice  and $\tau$ on the valley degree of freedom of the Dirac spinors $\psi$ and we have emphasized 'certain' because there is a position dependent rotation of the the valley spin around $\tau_z$. { Note that the gradient contribution of the impurity is related to the  clean Dirac hamiltonian via inversion and the exchange of $\tau$ and $\sigma$, and therefore can be thought of as a locally chirality reversing effect.}  However, as for short range scatterers the { last constant} contribution is dominant  the effect should be negligible.  Indeed, the crucial point is that the gradient terms generate a  second contribution in the s-wave channel of the T-matrix:
\begin{align}
    T_{\boldsymbol{R}}(0) =  \frac{A 3\sqrt{3} t_2}{{t_2^2}/{t_c^2}-1} [\sigma_z \tau_z + \frac{t_2}{t_c}\sigma_x(\tau_x \cos f_{\boldsymbol{R}}+\tau_y\sin f_{\boldsymbol{R}} )] \nonumber 
\end{align}
where $f_{\boldsymbol{R}}= 2 \boldsymbol{K} \cdot \boldsymbol{R}$, with $\boldsymbol{K}$ being the usual Dirac point, takes values in multiples of $\frac{2\pi}{3}$,  $t_c= \frac{2\sqrt{3}\pi}{3\sqrt{3}+2\pi} t \approx t$ and $A$ is the unit cell area. Here we have made explicit the position dependent rotation of the valley spin  first pointed out by Schelter \emph{et al.} \cite{Schelter2011}.
To first order in the impurity density, the self-energy of the configuration averaged Green's function is given by the configuration averaged T-matrix. Therefore, the valley off-diagonal terms cancel and at zero energy we obtain, without self-consistency,
\begin{align}
    \Sigma(0) &= \frac{\rho}{A} \langle  T_{\varhexagon}(0) \rangle  = \frac{\rho 3\sqrt{3} t_2}{{t_2^2}/{t_c^2}-1} \sigma_z \tau_z. \label{eq:efmass}
\end{align}

Remarkably,  expression Eq. \eqref{eq:efmass} accounts for all of the peculiar features of the phase diagram. First, we observe that for small $t_2$ the gap is linear in both $\rho$ and $t_2$. In particular \emph{any} finite density of these impurities is enough to open a topological gap with $C = \sgn{t_2}$. This is natural consequence of the critical system having single particle correlation that decay with the inverse of the distance.  Second, beyond the critical value $t_c$ the sign of the mass term switches and therefore so does the Chern number. Third, at $t_c$ the T-matrix has a pole. This signals the appearance of two degenerate zero energy bound states (ZBS) \footnote{The eigenvalues of $VG$ causing the divergence are twice degenerate, therefore there a two bound states} acting as a seed for the thermal metal. While the TM phase is beyond the T-matrix description, it is still clear that a dilute number of ZBS alone can not be responsible for a metallic phase, unless the spectral gap of the Dirac fermions is closed. Extending the calculation to energies away from the PH-symmetric point we obtain a closed analytical but lengthy expression which we show in the supplemental material (SM). At $t_2$ close but not equal to $t_c$ we get for energies \mbox{$t \gg E \gg 
    |t_2-t_c|$ \footnote{Note that when taking the full expression and completing $\gamma(0) = 0$ we do recover Eq. \ref{eq:efmass}. However, the energy and the $t_2$ limit do not coincide because the self-energy is discontinuous.} :}
\begin{align}
    \Sigma(E)  \approx \frac{\rho}{A} \left( \gamma^{-1}\sigma_{0} \tau_{0} + \frac{ A 3\sqrt{3}t_2}{{t_2^2}/{t_c^2}-1} \gamma^{-2}\sigma_{z} \tau_{z} \right ) \\  \text{with }  
    \gamma  = \lim_{r \to 0} G_0(E,r)-G_0(0,r)\approx \frac{1}{2\pi v^2} {E\log iE} .
\end{align}
  We observe that $\Sigma_{00}$ is divergent, however at the value of the discontinuity, i.e., $E = 0$, $\Sigma_{00}$ is pinned to $0$ by PH-symmetry. Therefore we can conclude that the spectral gap is indeed closed by an impurity resonance, the role of which will be clarified in the discussion following the numerical results. 
  
\emph{Numerical Results.}--- The Chebychev polynomial Green's function (CPGF) expansion is a formally exact rewriting of the Green's function of the rescaled Hamiltonian $h$ whose spectrum lies in the interval $[-1,1]$:
\begin{align}
    G(E+i \eta) = \sum_{n = 0}^\infty \lambda_n(E,\eta) P_n(h), \nonumber
\end{align}
where $\lambda_n$ are system independent coefficients and $T_n$ is the $n$th order  Chebyshev polynomial. Physical observables like the spectral function \mbox{ $A(E,k) = -1/\pi \Im \Tr G^r(E,k)$} or the Kubo-Bastin conductivities \cite{Bruno2001}
\begin{align}
    \sigma_{ij} = \frac{\hbar}{2 \pi} \int d\epsilon \Tr ( J_i\partial_\epsilon G^r J_j-J_j\partial_\epsilon G^a J_i)(G^a-G^r) \nonumber
\end{align}are expressed as traces of retarded $G^r$ ($\eta = i 0^+$) and advanced $G^a$ ($\eta = i 0^-$) Greens functions and computationally easy operators like the charge currents $J$. The sum of $T_n$ applied to a vector can be  computed very efficiently using the Chebyshev recursion relations. Because of the sparseness of $P_n(h)$ \footnote{Sparseness of $P_n(h)$ depends on the connectivity of the lattice and N} the trace over the whole Hilbert-space can be reduced to just one random vector with an error of  $O(N^{-\frac{1}{2}})$ \cite{Weisse2006}, $N$ being the total number of orbitals. Combining this with a cutoff for $n$ proportional to $\eta^{-1}$ we are able to carry out large scale simulations (up to $N = O(10^9)$,  $\eta = O(10^{-4})$ ) using the \emph{KITE} software package \cite{Joo2020} to calculate the  density of states (DOS) and longitudinal conductivities $\sigma_{xx}$. While the calculation of $\sigma_{xx}$ can be simplified to require only one expansion \cite{Ferreira15}, calculations of the Hall conductivity $\sigma_{xy}$  require a double Chebyshev expansion. Nevertheless, we are able to access reasonably large system sizes ($N = O(10^7)$,  $\eta = O(10^{-3})$)  using the Fast Fourier-Chebyshev approach \cite{decastro2023fast}.
\begin{figure}
    \centering
    \includegraphics[width=0.49\textwidth]{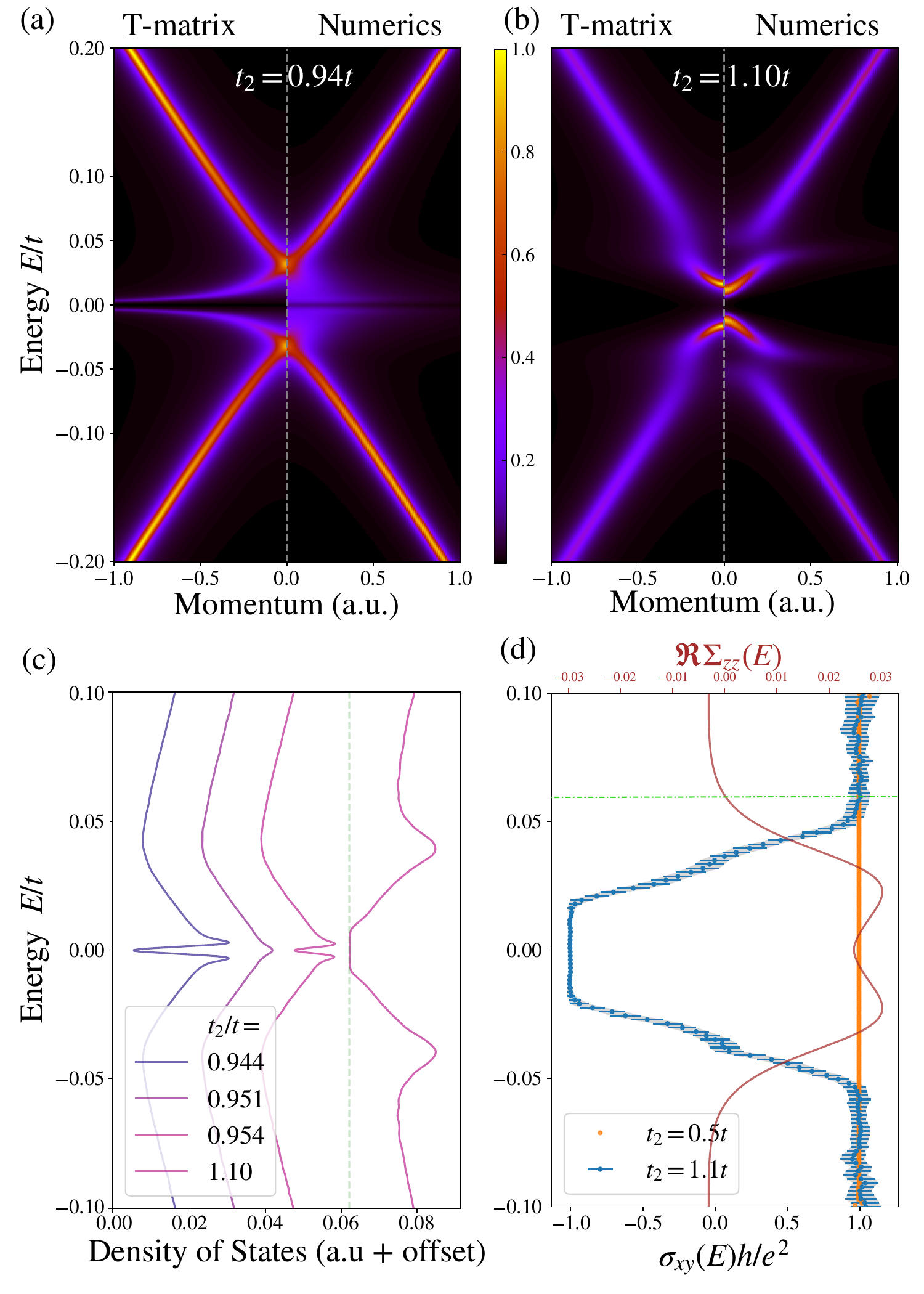}
    \caption{\textbf{(a)} $/$ \textbf{(b)} Spectral function along a momentum path through the Dirac point for $\rho = 0.15\%$ and $t_2 = 0.94/1.10$ rescaled to a maximum value of 1. The left half of each panel is computed analytically within the T-matrix approximation. \textbf{(c)} Density of states calculated using the CPGF  for various values of $t_2$. For better visibility the values have been off-set by a constant which is zero for the smallest $t_2$ and increased by $0.017$ (a.u.) for every increasing value of $t_2$. The vertical green dashed lines shows the origin of the rightmost curve. \textbf{(d)} Hall conductivity and it's standard deviation computed using the CPGF expansion averaged over 64 disorder configurations. We also plot the real part of the  self-energy mass $\Sigma_{zz}$ at $t_2 = 1.1$ whose values are indicated by the top scale. For reference, we also include the Hall conductivity data at $t_2 = 0.5$. The horizontal dashed line indicates the zero crossing of $\Re{\Sigma_{zz}}$. }
    
    \label{fig:mainfig}
\end{figure}

We estimated the phase boundaries in Fig.\ref{fig:enter-label} (a) by computing  
the periodic boundary version of the localization tensor \cite{Resta05}, and the Bott-index \cite{2010Hastings} for system sizes $N = O(10^3)$ and 40 disorder averages. The thermal metal phase can be identified by a diverging localization tensor and a non-quantized small value of the disorder-averaged Bott-index. We also calculated the Chern number for a point in the $C = -1$ phase using Fukui's method \cite{2005Fukui} on the Brioullin-Torus  of a super-cell. The obtained phase boundaries were verified  using  CPGF simulations for some exemplary points by scaling of the conductivity $\sigma_{xx}$ and a transition to logarithmically divergent DOS on the TM side \cite{Ivanov_2002}, further information can be found in the SM. We note that the transition to the thermal metal phase can not simply be determined by the onset of a diverging DOS at zero energy, as we observe that the insulating state may have a non-logarithmically divergent DOS. This divergence is consistent with the prediction of a Griffiths phase in symmetry class D where it appears as a result of rare impurity configurations with a large separation \cite{Mildenberger06}. The proposed mechanism is that zero modes bound to these impurities contribute to the DOS at a hybridization gap exponentially  small in the separation. The existence of a Griffiths phase in our model hence suggests that ZBS exist also at densities beyond the T-matrix description, where, however, they do not  correspond to a diverging topological mass term. \\

\emph{Chirality reversal at the tri-critical point.}--- Focusing now on $\rho \ll 1$ and $t_2 \approx 1$
we observe that we can identify $t_c$ by the coalescing of the particle and hole resonance at zero energy, Fig. \ref{fig:mainfig} (c). Furthermore, at finite $\rho$ the TM phase extends over a finite region of the parameter space (not visible for parameters of the plot).  The value at which the resonances meet is shifted slightly towards larger $t_2$ with increasing $\rho$ due to self-consistency effects on $t_c(\Sigma(E))$. For example at $\rho = 0.15\%$ it's around 0.950, while it is $0.955$ for $\rho = 0.5\%$. The spectral function close to $t_2 = t_c$ reveals an additional structure emanating from the tip of the gapped Dirac cone, Fig. \ref{fig:mainfig} (a). Within the T-matrix approximation this seems like a sharp mode becoming asymptotically flat while approaching the value of the effective spectral gap. On top of self-consistency effects, higher order impurity scattering effects change the picture slightly by broadening this band, however, scaling of the DOS with $\eta$ suggest that the spectral gap is maintained even including these processes, as further discussed in the SM.  With increasing $t_2$ this almost flat band eventually detaches from the Dirac cone creating a second mobility gap with $C = 1$ then curving further upward until the $C = 1$ gap is closed. Figure \ref{fig:mainfig} (b) shows the spectral function for such an intermediary value with a $C = 1$ gap at finite energy. Remarkably, this $C = 1$ gap implies that the resonance-hybridized  band carries a Chern number of $\pm 2$; We emphasize that due to the emergence of  the additional mobility gaps there is no corresponding clean Haldane model that  reproduces the whole topological information as the unit cell must have a minimum of 4 sites \footnote{Note that each individual ground state may still be adiabatically connected to one that commutes with the elementary translations of the honeycomb lattice.}. This highlights the essential role played by the impurity resonance which could not appear in a clean system justifying the use of the term genuine. 

We note that here we have the interesting situation that the sign of the effective mass term changes within the spectrum and thus the argument relating the topological mass of the Dirac fermions to the Hall conductivity, which is a Fermi-sea property,  may be invalid. The problem of evaluating the topology of a system with energy dependent self-energy  has been discussed in the context of the many-body Chern number \cite{Wang2012,Gavensky2023}, and the result, that the topology only depends on the renormalized Green function at chemical potential, is also applicable to our case. We give an independent confirmation of this by computing $\sigma_{xy}$ using the CPGF expansion, Fig. \ref{fig:mainfig} (d).\\

\emph{Discussion and Outlook.}--- 
While the thermal quantum Hall transition lies at zero mean mass $\mu = 0$ and zero mass variance  $\sigma^2 = 0$ with $\sigma^2$ being marginally irrelevant \cite{Ludwig94,Morimoto2015},  recent results \cite{Pan_2021, Wang21, Beenakker10} from field theory and numerics agree that the tri-critical point, with a transition to a thermal metal, is located at  $\mu = 0$ but a finite  $\sigma^2$. To reconcile this with the fact that the variance  $\sigma^2 \propto \rho(1-\rho) t_2^2$ of our impurity distribution is zero at zero density, we note that this apparent inconsistency may be resolved by estimating the mass variance from the renormalized impurity potential $T$ whose divergence at the tri-critical point can make $\sigma^2$ finite when taking the limit $\rho \rightarrow 0$. Another possible explanation is that in addition to mass disorder the impurities act as vortex disorder, via the introduction of effective $\pi$-fluxes, which can be large independently of $\rho$. Indeed this possibility has led Read and Green to conjecture that the thermal Hall transition is generically unstable to the thermal metal \cite{Read2000a}. However, as demonstrated in the SM, we were not able to identify such a vortex defect using the paradigm for the classification of defect Hamiltonians \cite{Teo10}.

The formation of a ZBS in the continuum as a result of a paremetric tuning of two resonances is reminiscient of the Friedrich-Wintgen mechanism~\cite{Friedrich85}. However, given that both resonances become infinitely sharp, we posit that  it is a distinct phenomenon;  namely,  that the particle-hole symmetry related Chern critical impurity resonances interfere destructively as they cross at $E=0$, thereby forming two bound states of opposite chirality. A  picture thus emerges where the continuum of extended states hosting the BIC is itself created by the impurity. 
 \\

To summarize, we have introduced an  example of a GTAI, where impurity resonances change the topology of the clean limit and may even introduce additional topological mobility gaps. Remarkably, these resonance-hybridized Chernful  bands  correspond to quasiparticles that have finite momentum support.
Furthermore, our model shows that Anderson tri-critical point of class D may arise at zero impurity density through a resonant impurity potential which can bind fermionic zero modes. 

We note that the necessary cancelling of the inter-valley contributions via the position dependent valley spin rotation is also possible for certain periodic arrangements of impurities and the GTAI, like the TAI,  is \emph{not} necessarily an Anderson insulator in the sense that it does not depend on localization effects \footnote{This is not true for the emergence of new topological mobility gaps which do depend on Anderson localization}. Therefore, the GTAI could also be observed for certain superlattice potentials.

We believe that our model is relevant to a variety of physical systems, ranging from topological bosons, e.g in photonics and magnonics, to the amorphous Kitaev model. In photonics, for example, the Haldane model can be realized using gyromagnetic scattering centers to open a topological gap, and individual control of the local magnetic field has already been demonstrated \cite{Wang2009,Zhou2020, liu2020topological}. Similar control can also be achieved in other proposals, both photonic and electronic,  where the topological mass term can be realized through scattering centers in artificial graphene \cite{Gibertini1009, Lannebere2018,Lannebere2019}. In magnonics our model could be realized by impurities, e.g. chiral adsorbed molecules~\cite{alhyder2023achiral}, inducing a density of local Dzyaloshinskii–Moriya interactions in honeycomb ferromagnets. We would also like to point out that the impurity potential studied  in this work has formal similarity with the problem of adatom absorption in graphene leading to Kane-Mele mass impurities \cite{Weeks20111, Aires2018}. While in the Kane-Mele version one can not observe a GTAI, our analysis provides two important insights that should still apply. Firstly, a leading order expansion of the impurity potential around the Dirac points \cite{Aires2016} is not sufficient to get accurate estimates of the gap for strong impurities. Second, we find a natural explanation for the cancelling of inter-valley scattering effects via random absorption \cite{Jiang2012} and can even predict periodic configurations where inter-valley scattering is suppressed.

Our study is also linked to recent works on the amorphous Kitaev model where the presence of odd-plaquettes opens a topological gap towards a Chiral QSL \cite{Cassella2023,Grushin23}. Taking into account  that the effective three-spin term generated by a magnetic field may also be obtained perturbatively by integrating out the triangles of the Yao-Kivelson model \cite{Dusuel2008, Yao2007}, we argue that for moderate amounts of amorphous disorder, the odd-plaquettes can be considered as generating a local (Majorana) topological mass term of the type considered here, which  is consistent with the observed linear dependence between the gap and the odd-plaquette density \cite{Grushin23}.

The mechanism of  `local TR-breaking' impurities opening a global topological gap may also be relevant to the emerging field of chiral imprinting by adsorption of chiral molecules
~\cite{ray1999asymmetric,qian2022chiral,wan2302signatures} and the stability of the Dirac cones at the surface of topological insulators or superconductors to, e.g., magnetic impurities~\cite{wray_2010, katmis2016high}.
Beyond the realization in concrete systems, we hope this work will also inspire future studies into Anderson criticality in class $D$, as well as this new type of ZBS. Furthermore,  our method of creating a topological gap  may also prove useful to further exploration of  the criticality of Dirac fermions in symmetry class $A$ \cite{Sbierski21},  because the small coordination number allows for very efficient CPGF calculations which rely on the sparseness of $P_n$.

\emph{Acknowledgements}--- We thank Santiago Gimenez de Castro for providing us with his implementation of the FastCheb algorithm and Aires Ferreira, João Manuel Viana Parente Lopes, and Simão Meneses João for providing support with QuantumKite. Furthermore, we acknowledge helpful discussions with Jonas Habel, Alaric Sanders, Wojciech Jankowski and Rao Peng. We also thank Aires Ferreira for providing thoughtful feedback and pointing us to the literature of adatom absorption in graphene.

We acknowledge support from the Imperial-TUM flagship partnership, the Deutsche Forschungsgemeinschaft (DFG, German Research Foundation) under Germany's Excellence Strategy--EXC--2111--390814868, DFG grants No. KN1254/1-2, KN1254/2-1, and TRR 360 - 492547816 and from the International Centre for Theoretical Sciences (ICTS) for the program "Frustrated Metals and Insulators" (code: ICTS/frumi2022/9), as well as the Munich Quantum Valley, which is supported by the Bavarian state government with funds from the Hightech Agenda Bayern Plus. F. P. acknowledges support from European Union’s Horizon 2020 research and innovation program under grant agreement No. 771537.

\providecommand{\noopsort}[1]{}\providecommand{\singleletter}[1]{#1}%

\ifarXiv
    \foreach \x in {1,...,\numbersupplementpages}
    {
        \clearpage
        \includepdf[pages={\x,{}}]{\supplementfilename}
    }
\fi

\end{document}